\documentclass[twocolumn]{aastex62}

\usepackage{apjfonts}

\shorttitle{More Nebulae around Nova-like Variables}
\shortauthors{Bond et al.}

\newcommand{\Ha}{H$\alpha$}
\newcommand{\Hb}{H$\beta$}

\newcommand{\kms}{{\>\rm km\>s^{-1}}}

\def\hei{\ion{He}{1}}
\def\heii{\ion{He}{2}}

\def\nii{\ion{N}{2}}

\def\oiii{\ion{O}{3}}

\def\Gaia{{\it Gaia}}
\newcommand{\GALEX}{{\it GALEX}}

\def\WISE{{\it WISE}}

\newcommand{\TESS}{{\it TESS}}

\begin{document}

\title{Two More Bow Shocks and Off-Center \Ha\ Nebulae Associated with Nova-like Cataclysmic Variables\footnote{Based in part on observations obtained with the Hobby-Eberly Telescope (HET), which is a joint project of the University of Texas at Austin, the Pennsylvania State University, Ludwig-Maximillians-Universit\"at M\"unchen, and Georg-August Universit\"at G\"ottingen. The HET is named in honor of its principal benefactors, William P. Hobby and Robert E. Eberly.} }

\author[0000-0003-1377-7145]{Howard E. Bond}
\affil{Department of Astronomy \& Astrophysics, Penn State University, University Park, PA 16802, USA}
\affil{Space Telescope Science Institute, 
3700 San Martin Dr.,
Baltimore, MD 21218, USA}

\author[0009-0008-5193-4053]{Calvin Carter}
\affil{Rocket Girls Ranch Observatory,
7215 Paldao Dr.,
Dallas, TX 75240, USA }

\author{Eric Coles}
\affil{Sierra Remote Observatories,
42130 Bald Mountain Rd., 
Auberry, CA  93602, USA}


\author[0009-0005-3715-4374]{Peter Goodhew}
\affil{Deep Space Imaging Network, 108 Sutton Court Rd., London, W4 3EQ, UK}

\author[0000-0002-9018-9606]{Dana Patchick}
\affil{Deep Sky Hunters Consortium, 1942 Butler Ave. Los Angeles, CA 90025, USA }

\author[0009-0009-3986-4336]{Jonathan Talbot}
\affil{Stark Bayou Observatory, 1013 Conely Cir., Ocean Springs, MS 39564, USA}

\author[0000-0003-2307-0629]{Gregory R. Zeimann}
\affil{Hobby-Eberly Telescope, University of Texas at Austin, Austin, TX 78712, USA}

\correspondingauthor{Howard E. Bond}
\email{heb11@psu.edu}

\begin{abstract}

We report discoveries of bow-shock nebulae, seen in \Ha\ and [\oiii] $\lambda$5007, around two cataclysmic variables (CVs): LS~Pegasi and ASASSN-V J205457.73+515731.9 (hereafter ASASJ2054). Additionally, both stars lie near the edges of faint extended \Ha-emitting nebulae. The orientations of the bow shocks are consistent with the directions of the objects' proper motions. The properties of LS~Peg and ASASJ2054, and of their nebulae, are remarkably similar to those of SY~Cancri, which we described in a recent paper; SY~Cnc is a CV likewise associated with a bow shock and an off-center \Ha\ nebula. These objects join V341~Arae and BZ~Camelopardalis, CVs that are also accompanied by similar nebulae. All five stars belong to the nova-like variable (NLV) subclass of CVs, characterized by luminous optically thick accretion disks that launch fast winds into the surrounding space. We suggest that the bow shocks and nebulae result from chance encounters of the NLVs with interstellar gas clouds, with the stars leaving in their wakes Str\"omgren zones that are recombining after being photoionized by the CVs' ultraviolet and X-ray radiation. Our discoveries illustrate the power of small telescopes equipped with modern instrumentation, and used to accumulate extremely long exposure times, for the detection of very low-surface-brightness nebulae.

\null\vskip 0.2in

\end{abstract}



\section{Introduction: Faint Nebulae Associated with Cataclysmic Variable Stars}

Cataclysmic variables (CVs) are short-period binaries in which a Roche-lobe--filling star transfers mass to a white-dwarf (WD) companion. Except in cases where the WD has a strong magnetic field---the AM~Herculis subclass---the transferred gas forms an accretion disk, from which most of the material eventually falls onto the WD\null. However, a small portion of the gas can be driven into space through a fast wind launched from the disk. The main subclasses of CVs are  classical novae (CNe, in which the accreted hydrogen ignites nuclear fusion on the surface of the WD); dwarf novae (DNe, in which the accretion disk is usually optically thin, but becomes optically thick and brighter during occasional outbursts); and nova-like variables (NLVs, in which the transfer rate is so high that the accretion disk remains optically thick most or all of the time). For comprehensive reviews of CVs, see the monographs and papers by \citet{Warner1995}, \citet{Szkody2012}, and \citet{Sion2023}. 

In the case of CNe, ejecta from their eruptions form expanding nebulae, which are typically visible for decades to a century or more, before dissipating into the interstellar medium (ISM). A few cases of very faint nebulae centered on NLVs have been discovered, often showing a hollow and filamentary structure. These are plausibly attributed to ejection from CN outbursts of the central binaries that occurred several centuries ago, but went unobserved. Aside from these nova shells, nebulae around CVs are remarkably rare, in spite of several extensive searches. We reviewed the literature on this subject in a recent paper \citep[][hereafter Paper~I]{BondSYCnc2024}, which gives further details and references.

In Paper~I we reported the serendipitous discovery of a very low-surface-brightness nebula associated with the NLV SY~Cancri. The discovery, and follow-up deep imaging, was made by co-authors of Paper~I who are advanced amateur astronomers. They work with small telescopes having fast focal ratios, narrow-band emission-line filters, and CMOS cameras. This equipment is used to accumulate long exposure times over many nights. 

The SY~Cnc nebula exhibits a bow-shock morphology, arising from the collision between a fast wind from the binary's accretion disk and the surrounding ISM\null. The bow shock lies on the front side of the star's path through the ISM, based on the known direction of its proper motion. Additionally, SY~Cnc is located at the front edge of a larger and fainter \Ha-emitting nebula. These features make SY~Cnc remarkably similar to another NLV, V341~Ara, which is also accompanied by a bow shock and lies well off-center in a faint \Ha\ nebula; see Paper~I for discussion and references.

In this paper, we present a further two objects that we have found to belong to this class of NLVs lying at the edges of faint \Ha\ nebulae, and accompanied by bow-shock nebulae as they plow their way through the ISM\null. We describe the discoveries of these objects, and our follow-up spectroscopy and deep imaging. We then close by briefly discussing possible scenarios for the origin of these phenomena, and by suggesting several useful follow-up investigations.

\section{LS Pegasi (S 193)}

\subsection{Photometric and Spectroscopic Behavior \label{sec:LSPeg_properties} }

LS~Peg is a relatively bright CV that has been recognized for nearly four decades and has an extensive history of multi-wavelength studies. Its variability was discovered by \citet{Hoffmeister1935}, who considered it a possible Mira variable.  The second column of our Table~\ref{tab:DR3data} lists astrometry and photometry of the star from \Gaia\/ Data Release~3\footnote{\url{https://vizier.cds.unistra.fr/viz-bin/VizieR-3?-source=I/355/gaiadr3}} (DR3; \citealt{Gaia2016, Gaia2023}). The bottom row gives its nominal absolute magnitude in the \Gaia\/ system, determined from its trigonometric distance of $\sim$290~pc.\footnote{A Bayesian analysis of {\it Gaia\/} EDR3 data by \citet{BailerJones2021} gives a distance of $286.3^{+3.1}_{-3.2}$~pc.} Here we have adopted an interstellar reddening of $E(B-V)=0.03$ at the distance of the star, obtained using the online {\tt GALExtin} tool.\footnote{\citet{Amores2021}; \url{http://www.galextin.org/}}


\begin{deluxetable}{lcc}[h]
\tablecaption{\Gaia\/ DR3 Data for LS Peg and ASASJ2054 \label{tab:DR3data} }
\tablehead{
\colhead{Parameter}
&\colhead{LS Peg}
&\colhead{ASASJ2054}
}
\decimals
\startdata
RA (J2000)  & 21 51 57.937 & 20 54 57.725   \\
Dec (J2000) & +14 06 53.27 & +51 57 31.92   \\
$l$ [deg] & 70.72 &  90.89  \\
$b$ [deg] & $-30.03$& +04.45  \\
Parallax [mas] & $3.430 \pm 0.039$ &  $1.795 \pm 0.012$  \\
$\mu_\alpha$ [mas\,yr$^{-1}$] & $19.641 \pm 0.036$ &  $-10.933 \pm 0.015$  \\
$\mu_\delta$ [mas\,yr$^{-1}$] & $-14.890 \pm 0.039$ &  $-2.945 \pm 0.014$ \\
$G$ [mag] & 11.89 &   13.67  \\
$G_{\rm BP}-G_{\rm RP}$ [mag] & 0.22 & 0.44  \\
$M_G$ [mag] & +4.5 & +4.6  \\
\enddata
\end{deluxetable}

The star attracted little attention over the half-century following its announcement, until it appeared in a catalog of new \Ha\ emission stars at Galactic latitudes above $10^\circ$ discovered by \citet{Stephenson1986} on objective-prism plate material. Stephenson remarked that the object\footnote{The star is no.~193 in Stephenson's catalog. Several ensuing publications designated the source as ``S\,193.'' Here we use the designation LS~Peg, assigned by \citet{Kazarovets1997}.} is ``emphatically neither M-type nor even red.''

This finding prompted a series of follow-ups. \citet{Downes1988} obtained an image-tube-scanner spectrum of the object, confirming its strong Balmer emission lines, and strongly suggesting that it is a CV\null. \citet{Garnavich1992} investigated the historical behavior of LS~Peg on nearly 1000 photographs from the Harvard plate collection covering more than 90~years. These data showed the star normally remaining bright at $B\simeq12$, but with occasional excursions---sometimes as rapidly as a few days---down to levels as much as $\sim$2~mag fainter. The star remains in such ``low states'' for durations of days to months. 

\begin{figure}[h]
\centering
\includegraphics[width=0.47\textwidth]{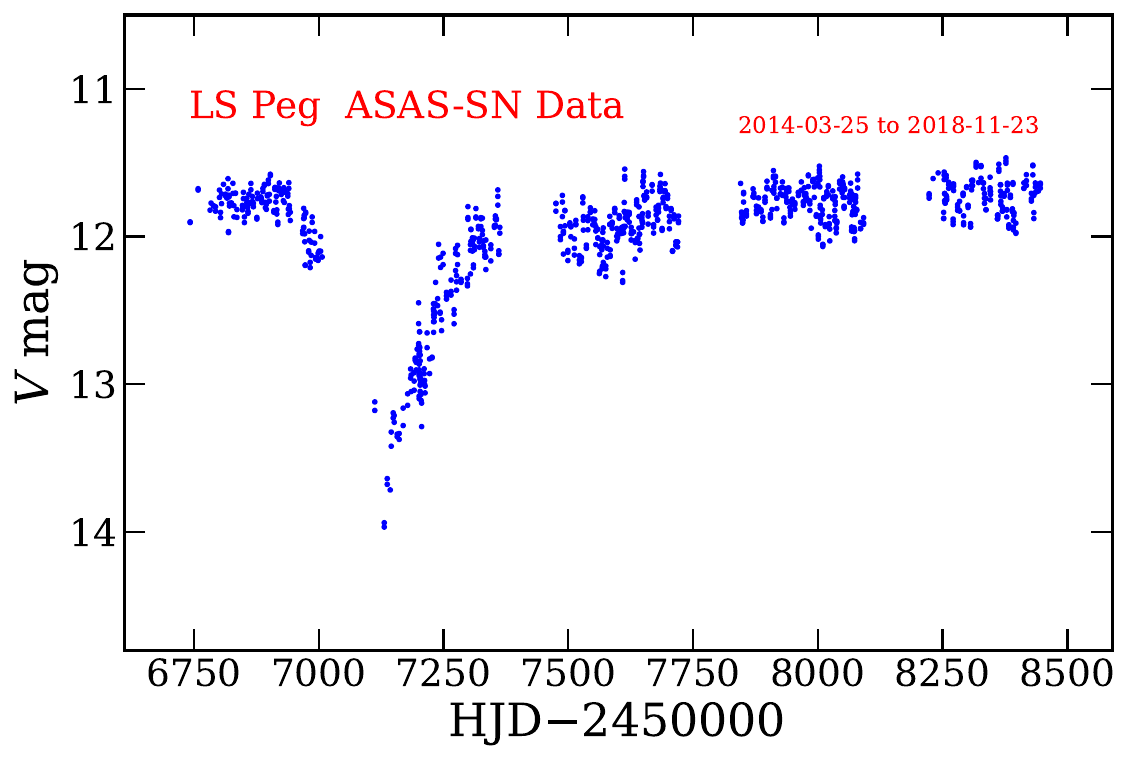}
\caption{
ASAS-SN $V$-band light curve of LS~Peg over a $\sim$4.5~year interval from early 2014 to late 2018. The star underwent a dramatic dimming episode, starting in 2014 October and lasting about 350~days. This confirms the star as a member of the VY~Scl subclass of nova-like cataclysmic variables. 
\label{fig:lspeg_lightcurve}
}
\end{figure}

We downloaded photometric data for LS~Peg from the website\footnote{\url{https://asas-sn.osu.edu}} of the All-Sky Automated Survey for SuperNovae \citep[ASAS-SN;][]{Shappee2014, Kochanek2017}. The ASAS-SN $g$-band data show no significant dimming events over an interval of about 7.3~years, from 2017 August to 2025 January. However, the earlier $V$-band observations reveal a single fading episode of more than 2~mag, with a duration of nearly one year, which can be seen in the light curve plotted in our Figure~\ref{fig:lspeg_lightcurve}. 

The photometric behavior of LS~Peg places it in the ``VY~Sculptoris'' subclass of CVs. VY~Scl stars are NLVs that are normally in ``high'' states, but which undergo occasional fading episodes due to transient reductions in the mass-transfer rate from the Roche-filling companions \citep[e.g.,][]{Honeycutt2004, Schmidtobreick2017}. For discussions of the VY~Scl subclass, see, for example, \citet{Warner1995}, \citet{Leach1999},  \citet{Sion2023}, and \citet{Duffy2024}.\footnote{The ``VY~Scl'' designation  for this class of CVs was first (to his knowledge) proposed by H.E.B. in a talk at the 1981 Santa Cruz Summer Workshop on Cataclysmic Variables and Related Systems.}

Since the exploratory work in the late 1980s and early 1990s, there has been a variety of investigations of LS~Peg, and we only summarize a few important results here, along with a few new findings.

\citet{Taylor1999} carried out a detailed spectroscopic and photometric study of the system. From radial velocities of its \Ha\ emission line, they determined an orbital period of $0.174774\pm0.000003$~days (4.19~hours). Photometry showed a quasi-periodic variation with a typical period of 20.7~min (which had been noted earlier by \citealt{Garnavich1992}), sometimes reaching a peak-to-peak amplitude as large as $\sim$0.2~mag. \citet{Taylor1999} placed LS~Peg in the ``SW Sextantis'' subgroup of NLVs; this class is defined by spectroscopic properties, including orbital periods in the range of about 3--4~hours (they make up a dominant fraction of CVs in this period range), single-peaked emission lines, high-excitation spectral features, and the brief appearance of absorption at the Balmer and \hei\ lines around orbital phase 0.5. For a discussion of the SW~Sex stars, see the review by \citet{Schmidtobreick2017}; she notes that the SW~Sex phenomenon is common among CVs with high mass-transfer rates, which create hot and luminous WD primaries. Many objects belong both to the SW~Sex spectroscopic class and the VY~Scl photometric class.


\citet{Rodriguez2001} discovered circular polarization in their investigation of LS~Peg. The polarization varies with a period of 29.6~minutes, suggesting that the WD is magnetic and rotates at that period. The possiblity that magnetic fields play a key role in VY~Scl variables has been discussed by several authors, including \citet{Hameury2002}.

\begin{figure}[h]
\centering
\includegraphics[width=0.47\textwidth]{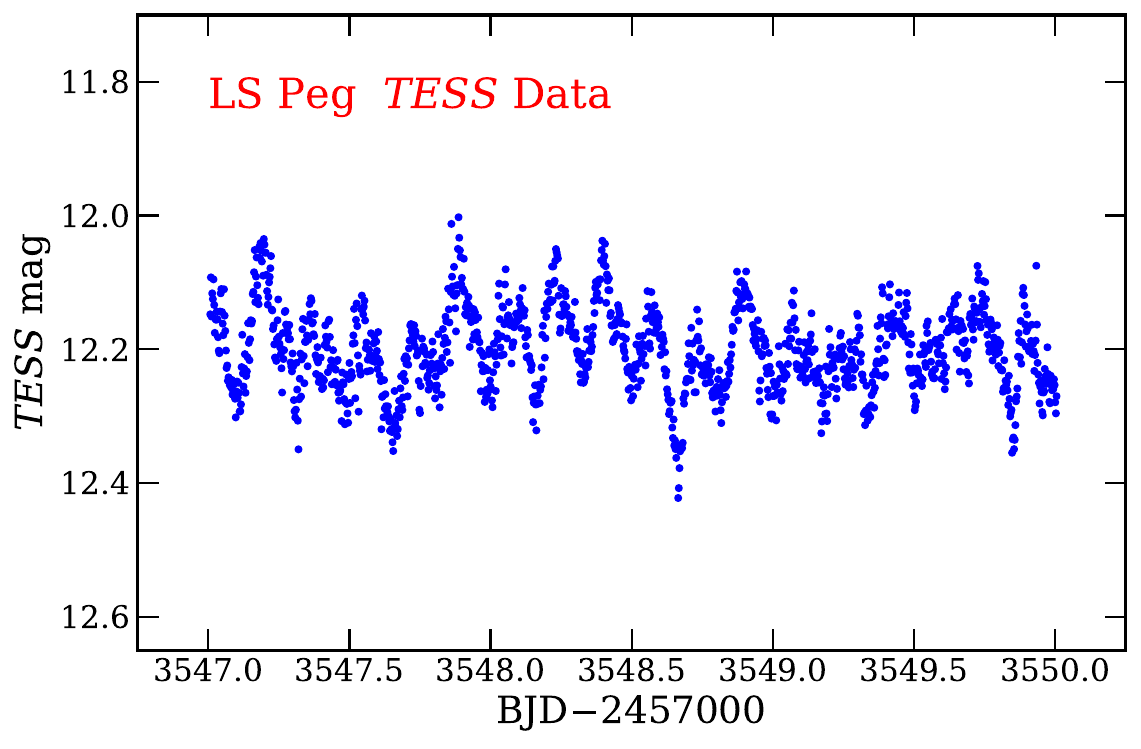}
\caption{
Representative \TESS\/ light curve for LS~Peg, covering a 3-day interval in 2024 August at a 200~s cadence. Variations are seen with a peak-to-peak amplitude of about 0.2~mag and a varying period typically around 0.170~day (4.08~hr), about 2\% shorter than the orbital period of 4.19~hr. 
\label{fig:lspeg_tess}
}
\end{figure}

We examined high-precision aperture photometry of LS~Peg obtained by the {\it Transiting Exoplanet Survey Satellite\/} (\TESS) mission, using the online {\tt TESSExtractor} tool.\footnote{\citet{Serna2021}; \url{https://www.tessextractor.app}}  Figure~\ref{fig:lspeg_tess} shows a sample of the \TESS\/ data, in this case excerpted from the Sector~82 run, having a cadence of 200~s. Measurements for an interval of 3~days are plotted, giving a representative view of the star's photometric behavior. Here we see a ``sawtooth'' light curve, with a peak-to-peak amplitude of about 0.2~mag, superposed on slower changes in baseline level.  A periodogram analysis shows that the variations generally have a timescale around 0.170~day (4.08~hr), but they are incoherent, i.e., the period wanders by about $\pm$0.002~day. The mean period is similar to the orbital period of 4.19~hr, but {\it less than\/} it by about 2\%. We saw a similar phenomenon in SY~Cnc, as discussed in Paper~I\null. These oscillations at periods a few percent shorter than the orbital period are called ``negative superhumps,'' which are interpreted as arising from a beat between the orbital period and a retrograde precession of a tilted accretion disk.  We did not see the $\sim$20.7-min quasi-periodic variations noted in earlier work (see above) in the \TESS\/ data that we examined, and a variation at the 29.6~min WD rotation period is also not present in the \TESS\/ data. As found in earlier studies, there are no eclipses.


The above discussion is only meant to highlight the complex and variable phenomena seen in LS~Peg; a detailed physical interpretation is beyond the scope or aims of this paper. We encourage further comprehensive studies.

\subsection{Discovery of Faint Nebulosity \label{subsec:lspeg_discovery} }

Co-author Patchick has successfully conducted searches for faint planetary nebulae (PNe) for many years, as a member of the Deep Sky Hunters Consortium (see, for example, \citealt{Jacoby2010} and \citealt{Ritter2023}). In 2020 August, in the course of examining wide-angle narrow-band \Ha\ images at high Galactic latitudes from 
the Virginia Tech Spectral-Line Survey\footnote{\citet{Dennison1998}; \url{http://www1.phys.vt.edu/~halpha/}} (VTSS), he noticed a faint nebula in Pegasus. It is also marginally visible in the red image of the field in the Space Telescope Science Institute Digitized Sky Survey\footnote{\url{https://archive.stsci.edu/cgi-bin/dss_form}} (DSS). 

A subsequent literature search revealed that the NLV LS~Peg lies within this nebulosity. This indicated that the nebula is unlikely to be a PN, and made it worthy of further investigation and deeper direct imaging. { The nebula has a very low surface brightness and is not detected in broad-band survey images in the far- and near-ultraviolet and near- and mid-infrared from the {\it Galaxy Evolution Explorer\/}\footnote{\GALEX\/ images can be obtained from SIMBAD at \url{https://simbad.cfa.harvard.edu/simbad/sim-fid}.} (\GALEX\/), the Two Micron All Sky Survey\footnote{\url{https://irsa.ipac.caltech.edu/applications/2MASS/IM/}} (2MASS), and the {\it Wide-field Infrared Survey Explorer\/}\footnote{\url{http://irsa.ipac.caltech.edu/frontpage}} (\WISE\/).}

\section{The Nova-like Variable ASASSN-V J205457.73+515731.9 }

\subsection{Photometric Behavior}

\begin{figure}
\centering
\includegraphics[width=0.47\textwidth]{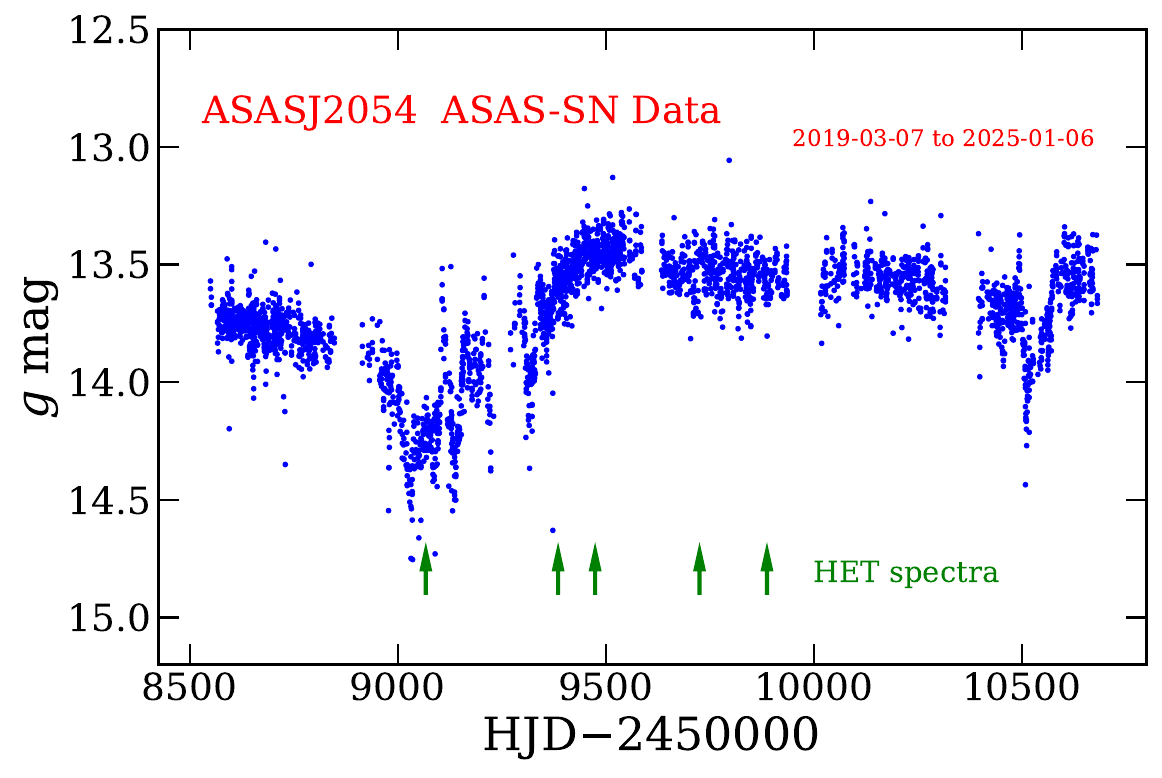}
\caption{
ASAS-SN $g$-band light curve of ASASJ2054 over a nearly six-year interval from early 2019 to early 2025. The star has shown several dimming episodes, making it a member of the VY~Scl class of cataclysmic variables. Green arrows mark the epochs of five spectroscopic observations with the HET, described in Section~\ref{subsec:asasj2054_spectroscopy}.
\label{fig:asasj2054_lightcurve}
}
\end{figure}

In contrast to the well-studied LS~Peg, ASASSN-V J205457.73+515731.9 (hereafter ASASJ2054) is a new variable star about which relatively little is known. Lying in Cygnus, it was found in the course of the ASAS-SN survey for transient and variable objects mentioned in Section~\ref{sec:LSPeg_properties}. On 2020 June~22, the ASAS-SN team \citep{Jayasinghe2020} called attention to their discovery that this object, normally at a magnitude of $g\simeq13.8$ ($V\simeq13.6$), had faded to about $g\simeq14.5$ over $\sim$20~days. Figure~\ref{fig:asasj2054_lightcurve} plots the $g$-band light curve of ASASJ2054 over a $\sim$6-year interval from early 2019 to early 2025, obtained by us from the ASAS-SN website. It shows that the mid-2020 fading lasted for about a year, with the object slowly returning to, or becoming even brighter than, its former brightness. However, the slow rise was interrupted several times by rapid fading events, with durations of roughly 10--15~days. More recently, in 2024 July, there was another such sharp fading episode, followed by a slow recovery to normal brightness taking about 75~days. \citet{Jayasinghe2020} also noted that a previous fading had been seen in ASAS-SN $V$-band data obtained in 2017, with a total duration of about 30~days. Based on its photometric behavior, these authors concluded that ASASJ2054 is a likely CV, belonging to the VY~Scl subclass. 


Table~\ref{tab:DR3data}, column~3, lists astrometric and photometric properties of ASASJ2054, taken from \Gaia\/ DR3. The \Gaia\/ parallax implies a distance of about 555~pc.\footnote{\citet{BailerJones2021} give a distance of $550.2^{+3.3}_{-4.3}$~pc.} Correcting for interstellar reddening at this distance of $E(B-V)=0.11$, according to the {\tt GALExtin} tool, we find an absolute magnitude of $M_G=+4.6$, as given in column~3 of the final row in Table~\ref{tab:DR3data}. This is similar to the absolute magnitude of LS~Peg, $M_G=+4.5$, and more generally is consistent with the typical absolute luminosities of other known NLVs; see, for instance, Paper~I, and the discussion below (Section~\ref{sec:discussion}). Because of their
bright, optically thick accretion disks in their normal high states, NLVs are 
considerably more luminous than dwarf novae at their normal minimum light. The latter are, roughly speaking, about 5~mag fainter than NLVs in their normal faint states.


\subsection{Association with Faint Nebulosity \label{subsec:asas_discovery} }

Immediately following the ASAS-SN team's announcement of their discovery of ASASJ2054, co-author Bond examined images of the object from the DSS\null. He found and reported \citep{BondASASSN2020} that ASASJ2054 is associated with a previously uncataloged very faint nebula, of approximate dimensions $3'\times4'$. This nebulosity is brightest in the red DSS image, suggesting that it is an \Ha\ emission nebula. The variable star is located near the western\footnote{\citet{BondASASSN2020} erroneously stated that the star lies near the eastern edge.} edge of the nebula, where the structure resembles that of a bow shock. 

These findings were quickly confirmed by \citet{Denisenko2020}. He posted images of the nebula online, showing renditions prepared both from the DSS frames\footnote{\url{http://scan.sai.msu.ru/~denis/J205457+515731-BRIR.jpg}} and from images\footnote{\url{http://scan.sai.msu.ru/~denis/J205457+515731-IPHAS.jpg}} obtained by the INT/WFC Photometric \Ha\ Survey of the Northern Galactic Plane \citep[IPHAS;][]{Barentsen2014}. This imagery show a faint, roughly elliptical nebula, with the variable star falling near its west-northwest edge.

{ As in the case of LS~Peg, this faint nebula is not detected in broad-band images from \GALEX, 2MASS, and \WISE.}

\subsection{Spectroscopy \label{subsec:asasj2054_spectroscopy} }

Prompted by these discoveries, and in order to investigate the nature of the star, H.E.B. obtained exploratory spectrograms of ASASJ2054  with the queue-scheduled (see \citealt{Shetrone2007PASP}) second-generation ``blue'' Low-Resolution Spectrograph (LRS2-B; \citealt{Chonis2016}) of the 10-m Hobby-Eberly Telescope (HET; \citealt{Ramsey1998,Hill2021}), located at McDonald Observatory in west Texas, USA\null. Briefly, LRS2-B employs a dichroic beamsplitter to send light simultaneously into two units: the ``UV'' channel (covering 3640--4645~\AA\ at a resolving power of 1910), and the ``Orange'' channel (covering 4635--6950~\AA\ at a resolving power of 1140). Data reduction is carried out by co-author Zeimann, using the \texttt{Panacea}\footnote{\url{https://github.com/grzeimann/Panacea}} and \texttt{LRS2Multi}\footnote{\url{https://github.com/grzeimann/LRS2Multi}} packages.
Further details of the LRS2-B spectrograph and data-reduction procedures are given, for example, in \citet{BondPaperI2023}.

The first LRS2-B spectrum of ASASJ2054, taken on 2020 August~6 and discussed briefly by \citet{Bond2020ASASJ2054spectrum}, showed a blue continuum and strong Balmer emission lines, fully consistent with the NLV classification. Eventually LRS2-B spectra were obtained at five epochs, as summarized in Table~\ref{tab:observations}. Green arrows at the bottom of the light curve in Figure~\ref{fig:asasj2054_lightcurve} mark the dates of the LRS2-B observations. The final column in the table gives the variable's $g$ magnitude from ASAS-SN at the date nearest that of the spectroscopic observation, usually within about 1--2~days or less.

\begin{deluxetable}{lccc}[h]
\tablecaption{Log of HET LRS2-B Observations of ASASJ2054\label{tab:observations} }
\tablehead{
\colhead{UT Date}
&\colhead{HJD$-$2450000}
&\colhead{Exposure}
&\colhead{$g$}\\
\colhead{[YYYY-MM-DD]}
&\colhead{}
&\colhead{[s]}
&\colhead{[mag]}
}
\startdata
2020-08-06 & 9067.884 & 90  & 14.17 \\
2021-06-20 & 9385.808 & 180 & 13.71 \\
2021-09-17 & 9474.777 & 360 & 13.46 \\
2022-05-26 & 9725.881 & 180 & 13.49 \\
2022-11-04 & 9887.640 & 180 & 13.53 \\
\enddata
\end{deluxetable}

\begin{figure*}
\centering
\includegraphics[width=0.9\textwidth]{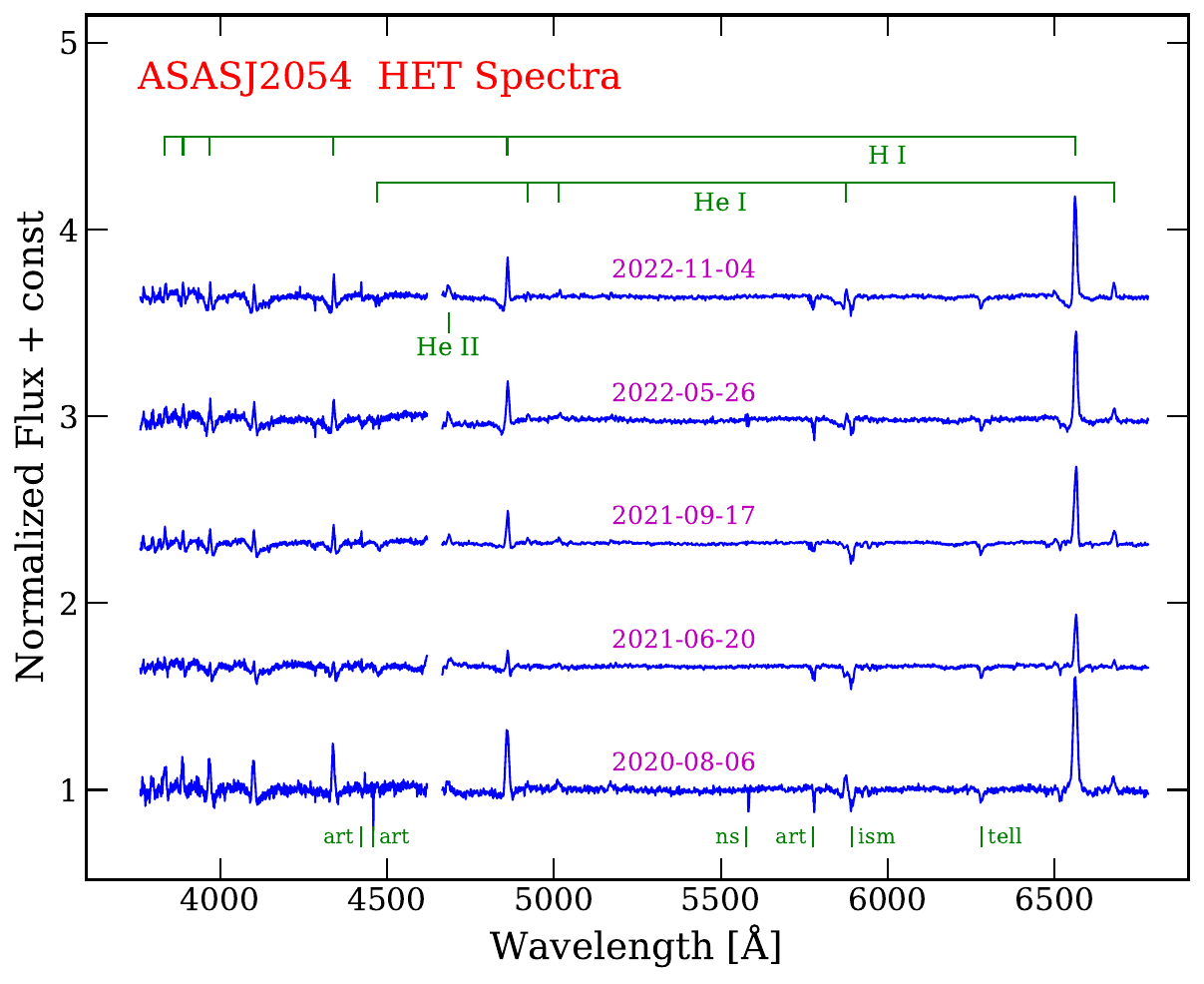}
\caption{
HET LRS2-B spectra of ASASJ2054 at the five epochs marked with arrows at the bottom of Figure~\ref{fig:asasj2054_lightcurve}. Spectra have been normalized to a flat continuum, and successive shifts by 0.66 of the continuum level have been applied to the observation sequence. Wavelengths of Balmer and \hei\--\heii\ lines are marked at the top. Features marked at the bottom are due to instrument artifacts (``art''), imperfectly subtracted night-sky emission (``ns''), and interstellar and telluric absorption (``ism'' and ``tell'').
\label{fig:asasj2054_spectra}
}
\end{figure*}

The LRS2-B spectra of ASASJ2054 are plotted in Figure~\ref{fig:asasj2054_spectra}. Here they are shown in order of the observation dates, starting with the earliest one at the bottom of the figure and proceeding upwards. The spectra have been normalized to a flat continuum level. After the first spectrum, the normalized fluxes at each date have been shifted upwards by successive steps of 0.66 of the continuum level.

Our first spectrum was obtained shortly after the deep light minimum in 2020, with the star still unusually faint at $g=14.17$. This spectrum exhibits strong emission lines of the Balmer series,\footnote{The LRS2-B integral-field units obtain spectra of the surrounding sky, which are subtracted from the stellar spectra in the reduction process. The nebulosity is extremely faint and is undetected in our relatively short HET exposures, except possibly for a very weak \Ha\ emission line.} and weaker emission lines of \hei\ and \heii. At the date of the next observation, almost a year later in 2021 June, ASASJ2054 had recovered to $g=13.71$, almost its normal high-state brightness. Now the emission lines are considerably weaker---a typical behavior for CVs in outburst, and NLVs in particular in their normal bright states. The final three spectra, later in 2021, and in mid- to late 2022, show the star in its high state. Absorption wings can be seen around the higher members of the Balmer series, again a typical feature associated with a bright, optically thick accretion disk. Of particular interest in the final two spectra are the apparent P~Cygni profiles at \Ha\ and \Hb, indicative of an outflowing wind from the accretion disk. Overall, the spectroscopic behavior of ASASJ2054 is remarkably similar to that of the VY~Scl system BZ~Cam \citep[see, e.g.,][their Figure~2]{Greiner2001}, including the appearance of P~Cygni features in the high-state spectra. As discussed below, and in Paper~I, BZ~Cam is also associated with a bow-shock nebula.

Relatively little is known at this stage about the exact nature of ASASJ2054---even its orbital period remains unknown (our five spectra are insufficient for this purpose). We examined the \TESS\/ data, but found no convincing evidence for periodic phenomena such as described above for LS~Peg; however, the source is relatively faint for \TESS\/ and the signal-to-noise ratio is fairly low. We see no evidence for eclipses, or even for an obvious orbital modulation, suggesting that the orbital inclination must be fairly low. We encourage detailed studies of this interesting object.

\clearpage

\section{Deep Imaging}

In this section we present and discuss deep images of the nebulosities associated with LS~Peg and ASASJ2054. The final imagery was created by combining large numbers of individual frames obtained by four (LS~Peg) and five (ASASJ2054) different telescopes feeding CMOS cameras. Technical details of the imaging equipment---optics, cameras, and filters---and of the reduction software are given in Table~\ref{tab:telescopes} in Appendix~A\null.

\subsection{LS Peg \label{subsec:lspeg_deep_imaging} }



Following up on Patchick's discovery (Section~\ref{subsec:lspeg_discovery}) of the faint nebulosity around LS~Peg, Goodhew, Talbot, and Carter obtained long-exposure images of the object. Goodhew used his dual-mounted 6-inch refractors located in Spain. His observations were carried out in 2020 August through December, using narrow-band \Ha\ and [\oiii] $\lambda$5007 filters for the nebulosity, and broad-band luminance and RGB bandpasses for the stellar field. Talbot employed both his 6-inch refractor in Mississippi and Carter's 13.8-inch reflector in Texas to obtain frames in narrow-band \Ha\  and broad-band RGB\null. These observations were carried out in 2024 September through early December. Details of the exposures are given in Table~\ref{tab:lspeg_exposures} in Appendix~A\null. The total exposure time across all filters and telescopes was 96.28~hr. 

\begin{figure*}
\centering
\includegraphics[width=6in]{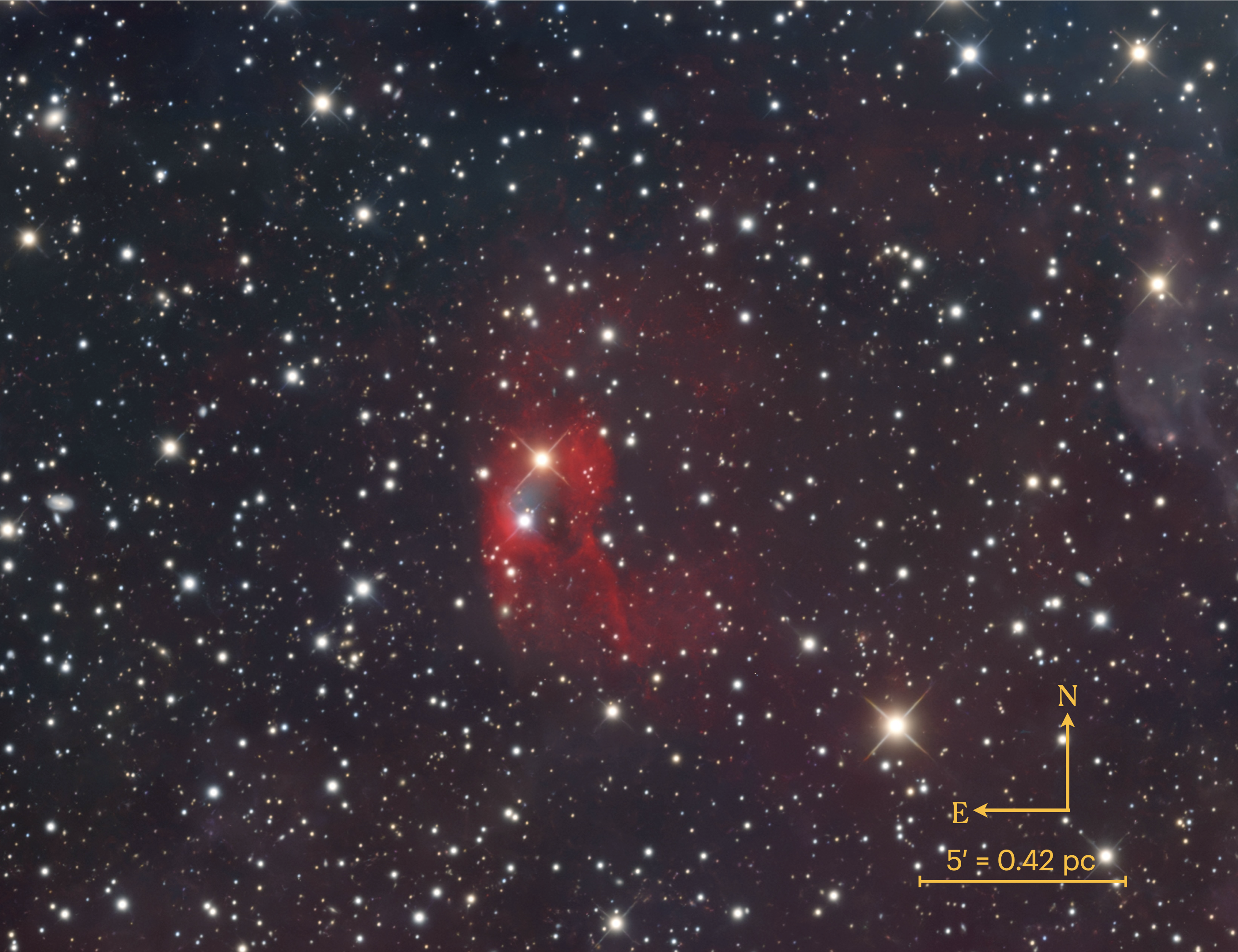}
\caption{Deep image of the nebulosity around the nova-like variable LS~Peg, from a total of 96.28~hours of exposure with four different telescopes (see Section~\ref{sec:lspeg_imaging_details} and Table~\ref{tab:lspeg_exposures} for details). Height of frame is $23'$. Luminance and RGB filters were used for the stellar background, and narrow-band frames in \Ha\ and [\oiii] $\lambda$5007 were mapped to red and blue-green, respectively. Orientation, angular scale, and linear scale at the distance of the variable star are indicated in the picture.  LS~Peg is identified in Figure~\ref{fig:lspeg_zoom}.
\label{fig:lspeg_image}
}
\end{figure*}

Figure~\ref{fig:lspeg_image} presents a color rendition of the resulting combined image, prepared by Talbot with input from Goodhew using the software listed in Section~\ref{sec:lspeg_imaging_details}. In this frame LS~Peg is seen to be located at the eastern edge of an extended \Ha-emitting\footnote{However, the \Ha\ 6562~\AA\ filter also has some transmission at the neighboring [\nii] emission lines at 6548 and 6583~\AA, which could be be contributing significantly.} nebula. The nebula has an approximately elliptical shape, elongated in the north-south direction, with diffuse edges and an apparently hollow interior. Its angular dimensions are very roughly $7'\times11'$ ($0.6\times0.9$~pc at the distance of the star). The overall structure of this emission nebula is remarkably similar to that of the one surrounding SY~Cnc, shown in Figures~1 and~2 in Paper~I\null. 

\begin{figure*}
\centering
\includegraphics[width=5in]{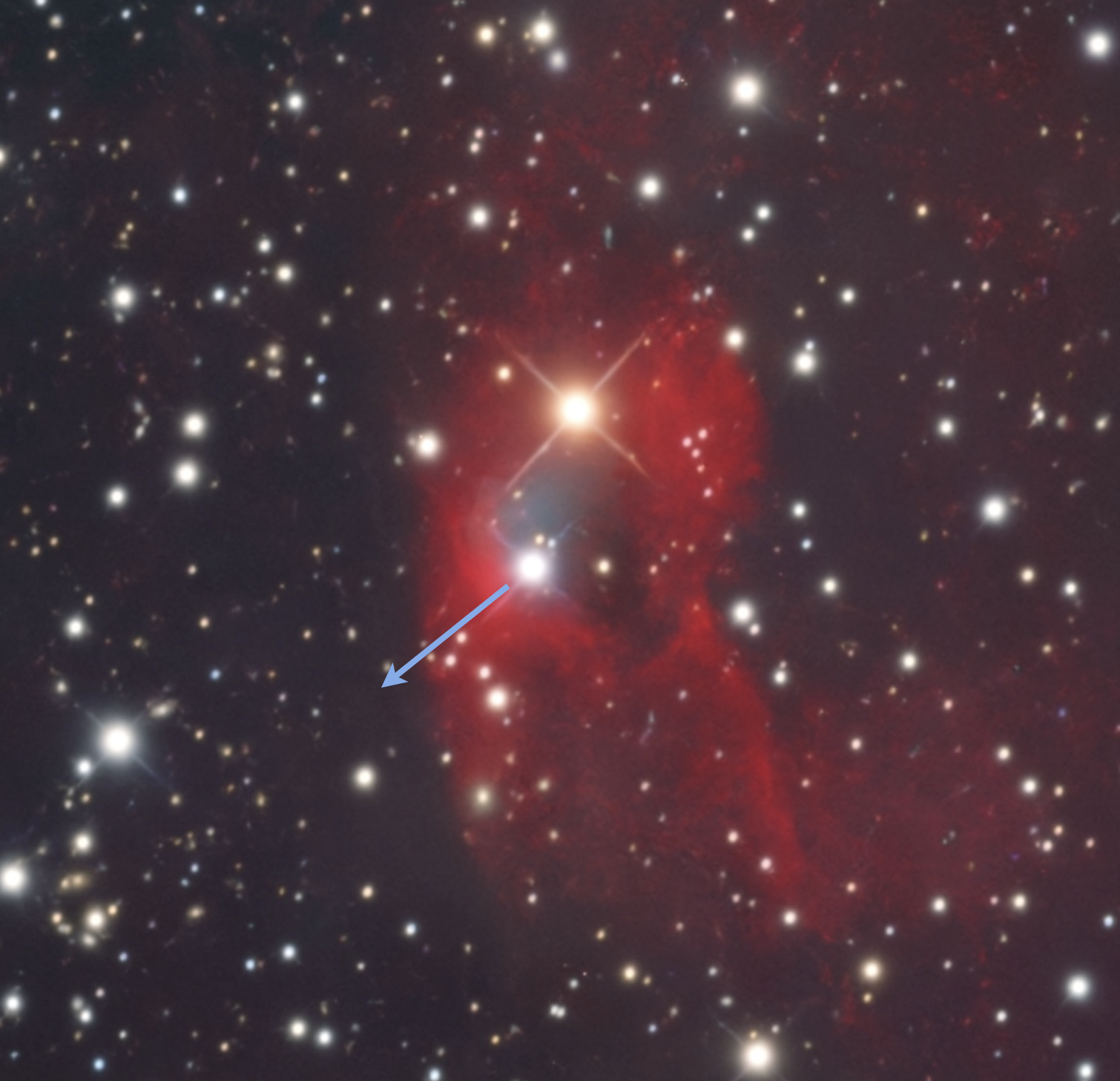}
\caption{Close-up of the LS~Peg nebula. The arrow is oriented in the direction of the proper motion of LS~Peg relative to the Galactic-rotation standard of rest at the star's location.
\label{fig:lspeg_zoom}
}
\end{figure*}

In Figure~\ref{fig:lspeg_zoom} we zoom in on the image of the LS~Peg nebula from Figure~\ref{fig:lspeg_image}. Here we see more clearly the morphology of a bow shock, in the form of a bright parabolic rim to the east of the star. This feature is prominent in the light of \Ha; the structure is also present in [\oiii], but is not shown very well because of the relatively short exposure time accumulated in [\oiii], only 5.91~hr versus 78.83~hr in \Ha.\footnote{We plan to obtain additional [\oiii] exposures later in 2025.} The bow-shock morphology is the signature of a fast wind from the star, colliding with the ISM as the star passes through it at high velocity. The blue arrow in Figure~\ref{fig:lspeg_zoom} lies in the direction of the \Gaia-based proper motion of the star, at north-through-east position angle $128\fdg7$; it shows that the  motion of the star is in a direction consistent with the interpretation as a bow shock.

The proper-motion direction depicted in Figure~\ref{fig:lspeg_zoom} has been adjusted (in this case by $+1\fdg6$) relative to the absolute \Gaia\/ value, in order to correct for the effect of differential Galactic rotation. The direction that is shown is thus made relative to the local standard of rest at the distance of the star. The correction formulae are presented by, among others, \citet{Moffat1998}, \citet{Comeron2007}, and \citet{Martinez2023}. For our calculations, we used a {\tt python} code\footnote{\url{https://github.com/santimda/intrinsic_proper_motion}. The file {\tt utils.py} at this site specifies the Oort constants and solar peculiar velocity used in the calculations, adopted from \citet{Wang2021}.} created by S.~del Palacio. The transverse velocity of the star, corrected for Galactic rotation using this code, is $22.0\pm0.4\,\kms$. The total space motion, relative to the standard of rest at the star's location, and based on a center-of-mass heliocentric radial velocity of $-56\pm14\,\kms$ \citep{Taylor1999}, is $55\pm14\,\kms$. 

\begin{figure*}
\centering
\includegraphics[width=6in]{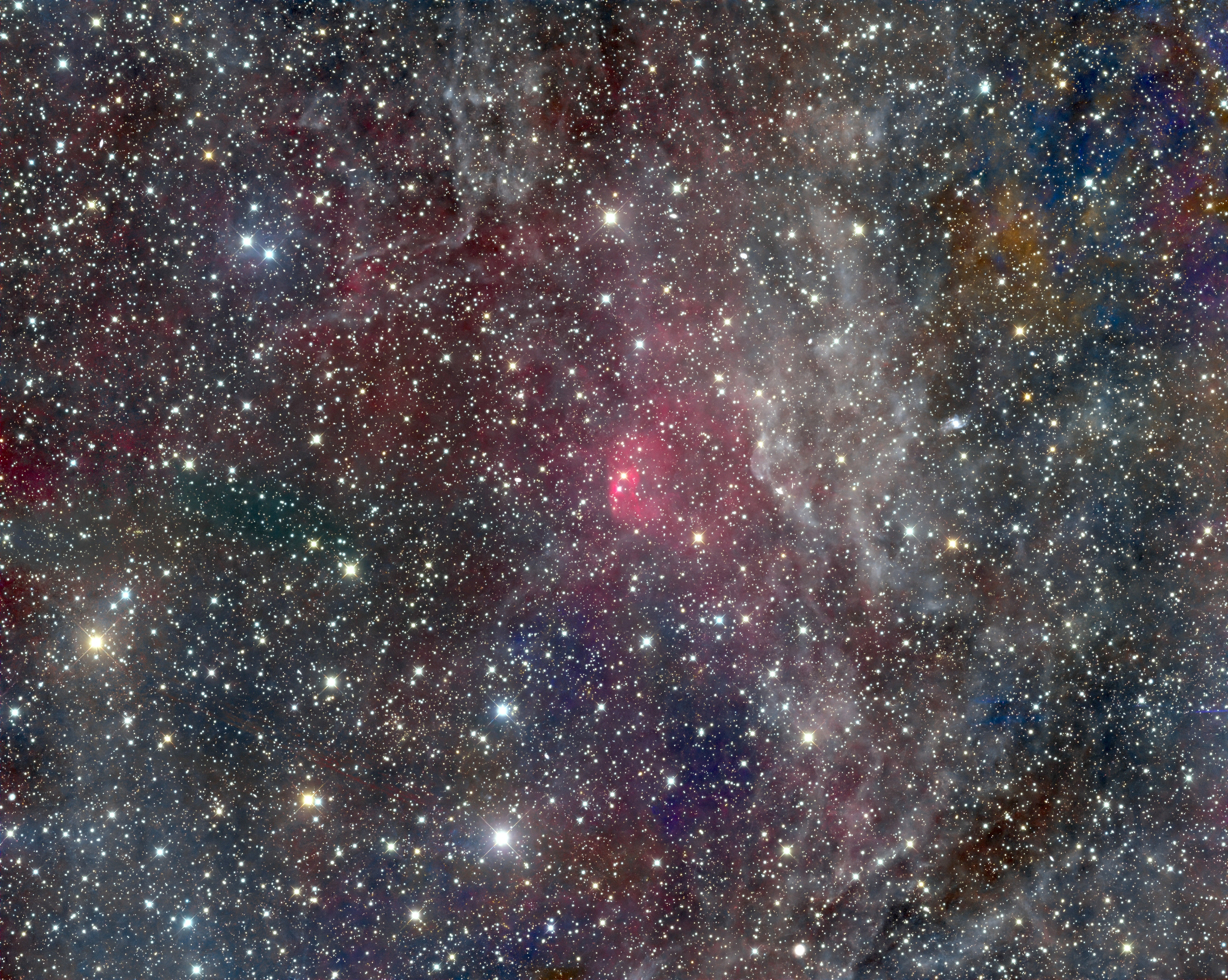}
\caption{Wide-angle deep image of the environment surrounding LS~Peg, emphasizing the broad-band filters to show reflection nebulosity in white. \Ha\ emission is colored pink. This frame's angular size is $2\fdg2\times1\fdg8$. 
\label{fig:lspeg_wideangle}
}
\end{figure*}

In Figure~\ref{fig:lspeg_image} we see faint patchy nebulosity well outside the \Ha\ emission nebula, especially in the western half of the frame. To illustrate the surrounding environment on a wider scale, Figure~\ref{fig:lspeg_wideangle} shows a frame with angular dimensions $2\fdg2\times1\fdg8$ (which is nearly the entire field of view of Carter's 13.8-inch reflector and CMOS camera). In this wide-angle rendition, the broad-band frames have been emphasized. The image shows that there is ambient reflection nebulosity covering most of the field, with the highest surface brightness in the western half. Emission in the narrow-band \Ha\ filter is brightest in the immediate vicinity of LS~Peg itself, but especially on its trailing (western) side. The morphology seen in the image is consistent with a scenario (see below) in which the LS~Peg \Ha\ emission nebula is excited by the passage of the variable star through clouds in the ambient ISM\null. The conspicuously off-center location of the star indicates that the nebulosity is unlikely to be material ejected by the star itself---for example, in a relatively recent (but unrecorded) nova outburst.

\clearpage

\subsection{ASASJ2054 \label{subsec:asas_deep_imaging} }

At the suggestion of Bond, deep images of the ASASJ2054 nebula were obtained by Coles, Goodhew, and Talbot, using the telescopes, filters, and exposure times listed in Table~\ref{tab:asas_exposures}. The grand total exposure time across all filters and telescopes was 144.33~hr, and the observations were obtained in 2024 August and September. 

\begin{figure*}
\centering
\includegraphics[width=6in]{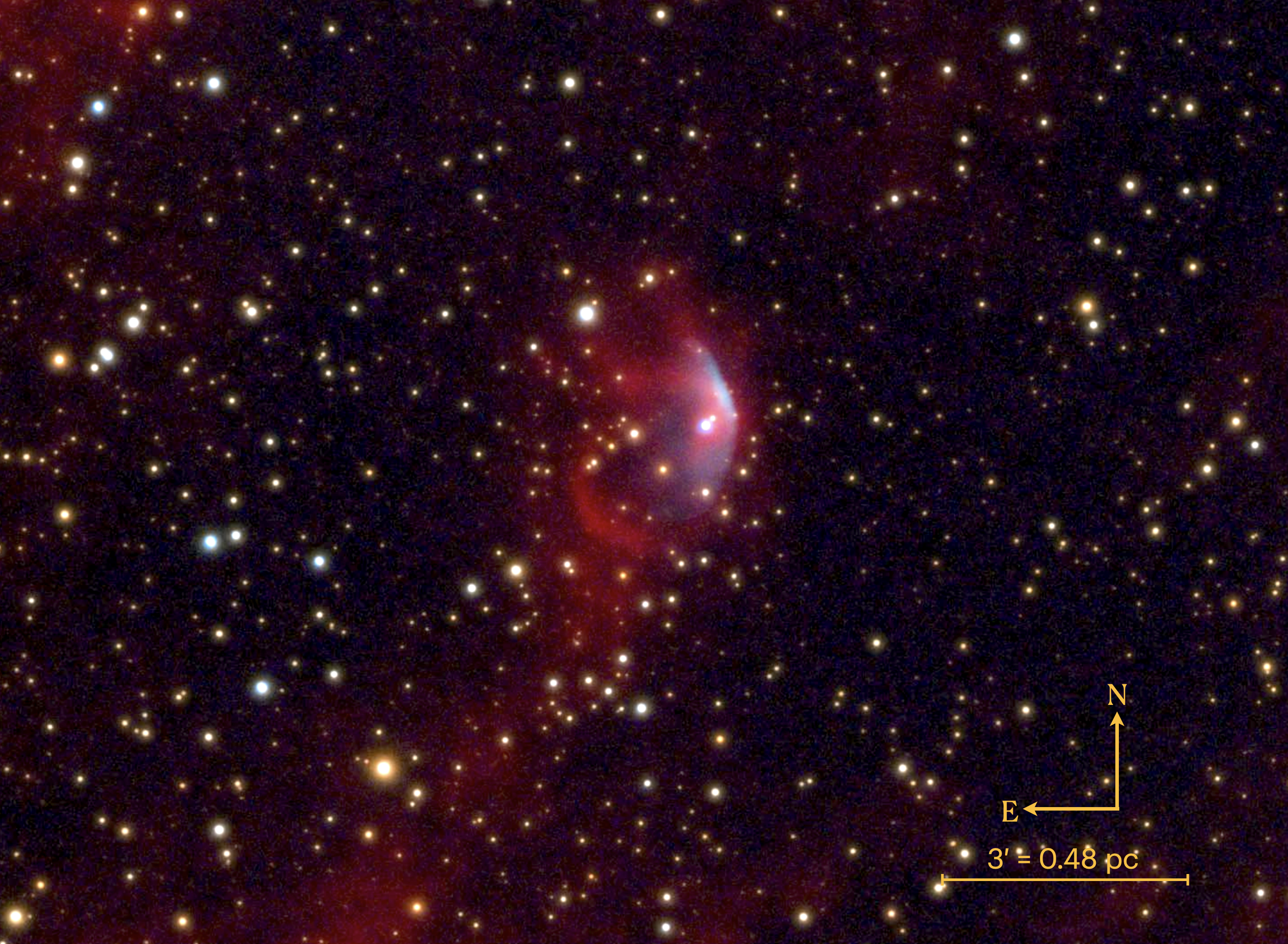}
\caption{Deep image of the nebula around the nova-like variable ASASJ2054, from a total of 144.33~hours of exposure with five different telescopes (see Section~\ref{sec:asas_imaging_details} and Table~\ref{tab:asas_exposures} for details). Height of frame is $11\farcm5$. RGB filters were used for the stellar field, and narrow-band frames in \Ha\ and [\oiii] $\lambda$5007 were mapped to red and blue-green, respectively. Orientation, angular scale, and linear scale at the distance of the variable star are indicated in the picture.  ASASJ2054 is identified in Figure~\ref{fig:asas_zoom}.
\label{fig:asas_image}
}
\end{figure*}

These frames were combined to create a color image by Goodhew, employing the software listed in Section~\ref{sec:asas_imaging_details}. 
The resulting rendition is shown in Figure~\ref{fig:asas_image}.\footnote{This image is also available online, with further details, at \url{https://app.astrobin.com/i/r85e3i}.} The nebula is, roughly speaking, a mirror image of the LS~Peg nebula presented in Figure~\ref{fig:lspeg_image}. As in the case of LS~Peg, the variable star lies well off-center, in this case falling on the western edge of an extended \Ha-emitting nebula. A parabolic bow shock lies to the west of the star.

The brightest parts of the nebula have an approximately elliptical shape, elongated in the north-south direction, with diffuse edges and an apparently hollow interior. Its angular dimensions are very roughly $3\farcm1\times3\farcm4$ ($0.55\times0.50$~pc at the distance of the star). Thus the linear size of the nebula, and its morphology, are remarkably similar to those for LS~Peg. One difference, however, is that it lies at the northern end of an extended wide filament of fainter \Ha\ emission, extending out of the frame to the south-southeast (see below for a wider-angle view of the surroundings).

\begin{figure*}
\centering
\includegraphics[width=5in]{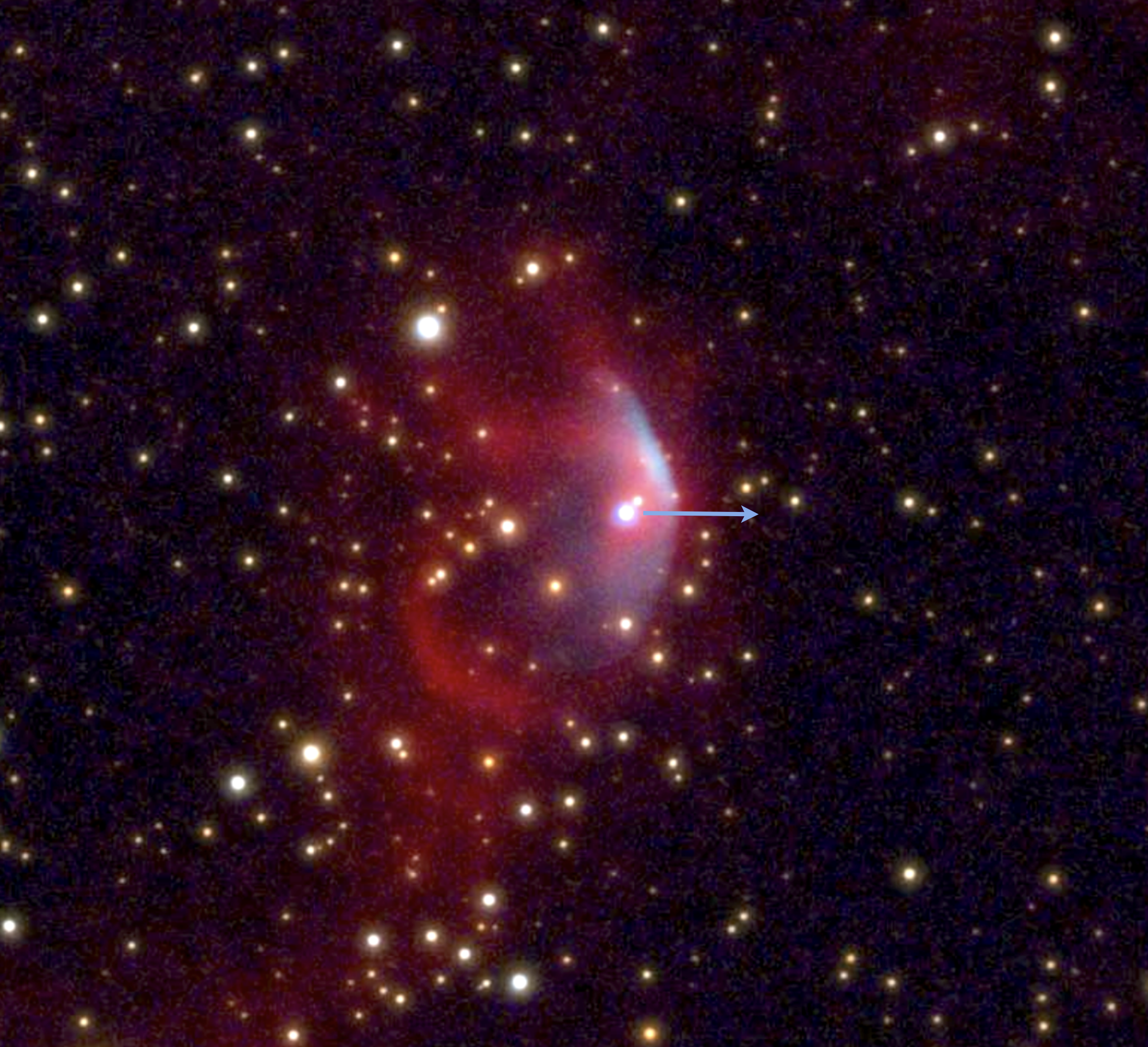}
\caption{Close-up of the ASASJ2054 nebula. The arrow is oriented in the direction of the proper motion of the nova-like variable relative to the Galactic-rotation standard of rest at the star's location.
\label{fig:asas_zoom}
}
\end{figure*}

Figure~\ref{fig:asas_zoom} zooms in on the ASASJ2054 nebula to show its structure more clearly. An arrow marks the direction of the star's proper motion from \Gaia\/ DR3,  corrected for Galactic rotation (by $+14\fdg5$) to show the direction of motion relative to the standard of rest at its location (using the software described in Section~\ref{subsec:lspeg_deep_imaging}). The bow shock clearly lies on the leading edge of the star's motion relative to the local medium. Its classical parabolic morphology is most clearly displayed in the light of [\oiii] $\lambda$5007. The transverse velocity of ASASJ2054, relative to the standard of rest, is $31.7\pm0.1\,\kms$, but its total space motion is unknown since we do not know its radial velocity.

\begin{figure*}
\centering
\includegraphics[width=6in]{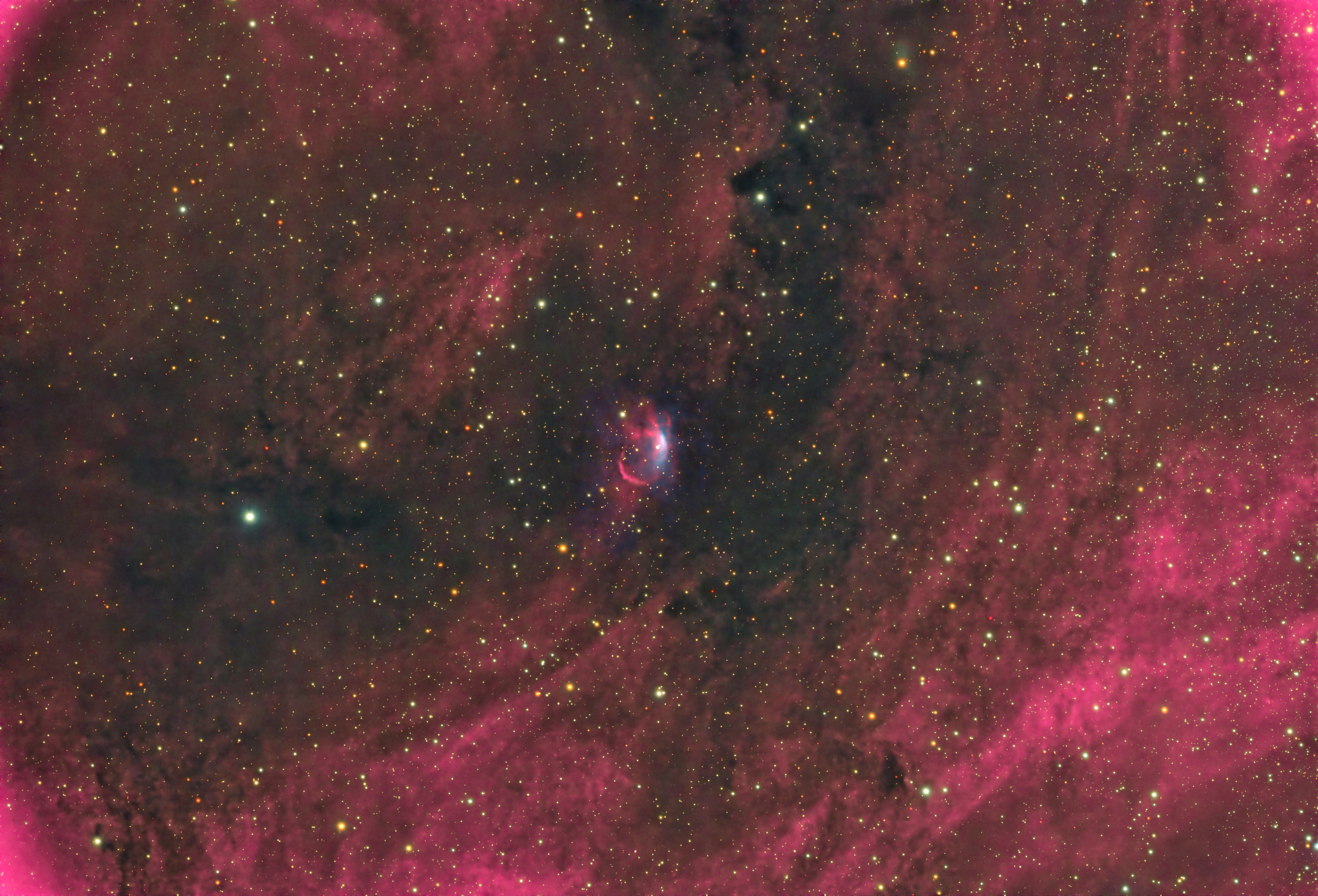}
\caption{A wide-angle view of the environment surrounding ASASJ2054. The angular dimensions of this frame are  $0\fdg90\times0\fdg61$. 
\label{fig:asas_wideangle}
}
\end{figure*}

As noted above, at the bottom left (south-southeast) side of the ASASJ2054 nebula in Figure~\ref{fig:asas_image} we see \Ha\ emission extending to the edge of the frame. Figure~\ref{fig:asas_wideangle} presents a wide-angle view of the surrounding field, prepared by co-author Coles. The rendition has been stretched to enhance the faintest nebular features. The \Ha\ filament seen in the previous two images extends for about a third of a degree; it appears to be part of a network of faint \Ha\ emission covering nearly the entire low-Galactic-latitude field. The overall appearance appears to be fully consistent with a scenario in which the nova-like star is passing through, and exciting, a region with a locally enhanced ISM density. { It is possible, however, that most of the wide-spread nebulosity in Figure~\ref{fig:asas_wideangle} lies considerably further away than the variable star.}


\clearpage

\section{Discussion \label{sec:discussion} }

Including the findings reported in this paper, we are aware of five cases in which a nova-like CV is associated with a bow shock seen in \Ha\ and [\oiii], with the star also lying conspicuously off-center within a larger extended faint \Ha-emitting nebula. In each instance, the displacement of the star from the center of the faint nebula is in the direction of its proper motion, and the bow shock is on the leading edge. These five objects are listed in Table~\ref{tab:list}, in order of discovery dates for the nebulae. Except for ASASJ2054, the nebula was discovered first, and only later was it realized that the nebula hosts a NLV\null. In the case of BZ~Cam, nothing was known about the star when the nebula was discovered by \citet{Ellis1984}, and subsequent observations revealed its NLV nature. V341~Ara was known to be a variable star, but it had been misclassified as a Cepheid (see \citealt{BondV341Ara2018} and references therein). SY~Cnc and LS~Peg were known NLVs before their nebulae were discovered, but, like the first two, the nebula discoveries did not result from a search targeted at CVs. Only the ASASJ2054 nebula was found as a result of a deliberate follow-up inspection of sky-survey images of a recently discovered NLV, as recounted in Section~\ref{subsec:asas_discovery}. 

\begin{deluxetable*}{lccccccc}[h]
\tablecaption{Nova-like Variables Associated with Bow Shocks and Off-Center \Ha\ Nebulae \label{tab:list} }
\tablehead{
\colhead{Nebula}
&\colhead{Star}
&\colhead{Distance}
&\colhead{$P_{\rm orb}$}
&\colhead{$G$}
&\colhead{$E(B-V)$}
&\colhead{$M_G$}
&\colhead{References\tablenotemark{a}}\\
\colhead{ }
&\colhead{ }
&\colhead{[kpc]}
&\colhead{[hr]}
&\colhead{[mag]}
&\colhead{[mag]}
&\colhead{[mag]}
&\colhead{}
}
\decimals
\startdata
EGB 4   & BZ Cam    & 0.37 & 3.69  & 12.97 & 0.11 & +4.8 & (1,2) \\ 
Fr 2-11 & V341Ara   & 0.16 & 3.65  & 10.81 & 0.02 & +4.8 & (3,4) \\ 
PaEl 1  & SY Cnc    & 0.41 & 9.18  & 12.69 & 0.04 & +4.5 & (5,6) \\ 
\dots   & ASASJ2054 & 0.55 & \dots & 13.67 & 0.11 & +4.6 & (7,8) \\ 
\dots   & LS Peg    & 0.29 & 4.19  & 11.89 & 0.03 & +4.5 & (9,10) \\ 
\enddata
\tablenotetext{a}{First reference for each star is for discovery of the nebula, and the second is for determination of the orbital period: (1)~\citet{Ellis1984}; (2)~\citet{Patterson1996}; (3)~\citet{Frew2006}; (4)~\citet{BondV341Ara2018, CastroSegura2021}; (5)~Paper~I \citep{BondSYCnc2024}; (6)~\citet{Casares2009}; (7)~\citet{BondASASSN2020}; (8)~Period unknown; (9)~Patchick, this paper; (10)~\citet{Taylor1999}.
} 
\end{deluxetable*}

For reviews of the substantial literature on stellar bow shocks see, for example, the introductory sections in recent papers by \citet{Kobulnicky2022} and \citet{Carretero2025}. Bow shocks are formed when a star with a fast stellar wind passes through the ISM at supersonic velocity. They are seen around a variety of objects, including OB stars, red supergiants, young stellar objects, and pulsars. The history given in the previous paragraph shows that the passage of a NLV through an ISM cloud is also a particularly favorable circumstance for the formation of a bow shock, along with an off-center surrounding faint nebula.

The first two columns in Table~\ref{tab:list} give designations\footnote{These designations were assigned in the literature using the standard nomenclature for PNe, i.e., abbreviated discoverers' names plus catalog numbers, before it was realized that the objects are actually not classical PNe.} for the NLV nebulae (except for the two new ones reported in the present paper), and for the associated variable stars. Column~3 give distances to the stars derived from their \Gaia\/ DR3 parallaxes, and the next column the orbital periods for the binaries (except for ASASJ2054, whose period remains unknown at this writing). The next three columns give the nominal apparent $G$ magnitudes from \Gaia\/ DR3, interstellar reddenings from {\tt GALExtin}, and the implied absolute magnitudes, $M_G$. 

The final column in Table~\ref{tab:list} gives literature references for the five objects; the first one for each object is for the discovery of the nebula, and the second is for a determination of the orbital period. The latter references are not intended to be complete, as several of the stars have been subjects of multiple investigations. As mentioned in our introduction, and discussed at greater length in Paper~I with literature references, a few other CVs not listed here are surrounded by, and roughly centered within, nebulae whose morphologies appear consistent with having been ejected from CN outbursts. As also summarized in Paper~I, a few CVs are known to be associated with bow shocks, but appear to lack the diffuse, off-center \Ha\ nebulae\footnote{For four of the objects in Table~\ref{tab:list}, the off-center \Ha\ nebulae are conspicuous in deep images. In the case of BZ~Cam, long-exposure images in [\oiii] and \Ha\ are dominated by its bow shock; however, deep exposures in \Ha, such as shown in Figure~4 in \citet{Greiner2001}, reveal a faint, diffuse, and roughly elliptical nebula. Relative to this nebulosity, BZ~Cam lies well off-center in the direction of its proper motion.} seen in the objects listed in Table~\ref{tab:list}.

The absolute magnitudes of these five stars cover a strikingly small range in $M_G$ of only +4.5 to +4.8. Absolute magnitudes this bright are typical of NLVs. They result from a relatively high rate of mass transfer from the donor star to the WD companion, of about $10^{-8}\,M_\odot\rm\,yr^{-1}$, which creates an optically thick and luminous accretion disk orbiting the WD component. As reviewed, for example, by \citet{Matthews2015}, not all of this material falls onto the WD: a bipolar outflow from the disk carries away some $\sim$1 to 10\% of the transferred matter into space as a fast wind. \citet{Matthews2015} (and references therein) note that the winds generally produce single-peaked Balmer emission lines in spectra of the binaries---of which those observed in LS~Peg \citep{Taylor1999} and in ASASJ2054 (our Figure~\ref{fig:asasj2054_spectra}) provide examples. For a recent discussion of winds from the accretion disks of nova-like CVs, see Section~8.1 of \citet{Matthews2025} and literature cited therein. 

These disk-launched outflows can be extremely fast. For example, in BZ~Cam, wind velocities as high as $3000\,\kms$ \citep{Kafka2004} and { a range of 4500 to} $8700\,\kms$ \citep{Balman2022} have been reported, based on X-ray, ultraviolet, and optical spectroscopy showing P~Cygni profiles. 


In our Paper~I, and in literature referenced therein, there are discussions of  scenarios that may account for the off-center locations of the NLVs with respect to the surrounding large and faint \Ha\ emission nebulae.
In one scenario (discussed in some detail for the case of V341~Ara by \citealt{CastroSegura2021}) the \Ha\ nebula is supposed to be material ejected from the NLV itself during an unobserved CN outburst in the past. In this picture, the extended nebula has to have been decelerated through its collision with the surrounding ISM, to explain why the star is now ``snowplowing'' through its own ejecta, as its fast wind creates the bow shock.  

An alternative scenario, which we favored in Paper~I, is that these objects represent chance encounters between a fast-moving NLV and a relatively stationary interstellar cloud---as is the case in most of the OB stars, pulsars, and other bow shocks discussed above. In addition to the bow shock, ultraviolet and X-ray radiation from the CV photoionizes the ISM as the star passes through it, creating the extended \Ha\ emission nebula. As the star moves on, the ISM slowly recombines, leaving a ``recombination wake'' behind the star---which explains its off-center location. Figure~\ref{fig:asas_wideangle} appears to be consistent with this picture, as it shows that the ASASJ2054 nebula lies at the northern end of an extended faint \Ha\ filament. This material clearly has not been ejected from the star, but is simply part of the ambient ISM through which the star is passing.  

A similar scenario for producing ionized nebulae around hot stars that have not ejected the material themselves is actually well-known in the PN field: there are many observed cases of hot subdwarfs producing photoionized Str\"omgren zones as they pass through interstellar clouds. In the PN literature these ionized nebulae have been called ``PN mimics,'' a terminology introduced by \citet{FrewParker2010}. It is common for the hot subdwarfs to lie off-center in these nebulae, as they leave behind a trail of recombining nebulosity. For a recent discussion of PN mimics, see \citet{Bond2023Fr2-30}, which presents a deep image of an PN mimic with an apparent recombination wake, obtained by co-author Talbot.





\section{Summary and Future Work}

In this paper we present deep broad-band and emission-line images of two NLVs---LS~Peg and ASASJ2054---obtained by advanced amateur astronomers using small telescopes and accumulating extremely long exposure times. These frames reveal bow shocks, along with faint extended \Ha\ nebulae in which the CVs are located well off-center. Our targets join three other NLVs that are likewise associated with off-center nebulae and bow shocks.

These objects are likely to be the results of chance collisions between NLVs and interstellar clouds. We use proper motions from \Gaia\/ to show that the bow shocks are on the leading edges of the stars' space motions, and we interpret the \Ha\ nebulae as being interstellar material photoionized by ultraviolet and X-ray emission from the NLVs. Our imagery shows recombination wakes on the trailing sides of the stellar trajectories. NLVs belong to a class of CVs that possess luminous accretion disks around their WD components, due to high mass-transfer rates from their donor stars. These accretion disks launch fast winds into the surrounding space, providing especially favorable circumstances for the formation of bow shocks. 

We suggest several avenues for future work. Deep imaging of a large sample of NLVs would be useful for a search for more bow shocks and interactions with the ISM\null. Amateurs are equipped to reach deeper surface-brightness levels than many existing surveys, as shown in this paper, and they have access to large amounts of observing time.

The orbital period of ASASJ2054 remains unknown, but could be determined relatively easily; it is likely to be approximately 4~hours. Ultraviolet spectroscopy of it and the other objects studied here would be of interest, as all of them are clearly driving substantial fast winds. Optical spectroscopy of the surrounding nebulae could characterize their electron temperatures and densities, which would help constrain scenarios for the origin of the nebulosity---but the nebulae are extremely faint.


\acknowledgments


This work uses data from the Virginia Tech Spectral-Line Survey (VTSS), which was supported by the National Science Foundation.


This work has made use of data from the European Space Agency (ESA) mission
{\it Gaia\/} (\url{https://www.cosmos.esa.int/gaia}), processed by the {\it Gaia\/}
Data Processing and Analysis Consortium (DPAC,
\url{https://www.cosmos.esa.int/web/gaia/dpac/consortium}). Funding for the DPAC
has been provided by national institutions, in particular the institutions
participating in the {\it Gaia\/} Multilateral Agreement.

The Digitized Sky Surveys were produced at the Space Telescope Science Institute under U.S. Government grant NAG W-2166. The images of these surveys are based on photographic data obtained using the Oschin Schmidt Telescope on Palomar Mountain and the UK Schmidt Telescope. The plates were processed into the present compressed digital form with the permission of these institutions. 



This research has made use of the SIMBAD and Vizier databases, operated at CDS, Strasbourg, France.

We acknowledge the Texas Advanced Computing Center (TACC) at The University of Texas at Austin for providing high-performance computing, visualization, and storage resources that have contributed to the results reported within this paper.

Preston Starr, Director of the Dark Sky Observatory, has been helpful in our work.

We thank S.~del Palacio, author of the {\tt python} code used in Sections~\ref{subsec:lspeg_deep_imaging} and \ref{subsec:asas_deep_imaging}, for pointing out to us the importance of correcting proper motions for the effect of Galactic rotation, and for kindly assisting us in implementing his code.

\clearpage

\appendix 

\section{Technical Details for the Deep Exposures}

Here we present details of the deep-imaging exposures for our targets LS~Peg and ASASJ2054. The telescopes, cameras, and filters that were used are summarized in Table~\ref{tab:telescopes}. Further information is available at the observers' websites: Coles: \url{https://www.astrobin.com/users/coles44}; Goodhew: \url{https://www.imagingdeepspace.com}; Talbot: \url{https://starscapeimaging.com}.

\begin{deluxetable}{lllll}[h]
\tablecaption{Telescopes, Cameras, and Filters \label{tab:telescopes} }
\tablehead{
\colhead{No.}
&\colhead{Observer}
&\colhead{Optics}
&\colhead{CMOS Camera}
&\colhead{Filters Used}
}
\startdata
1    & Coles & Planewave 20\,in $f/7.77$ reflector & Moravian C3-61000 Pro & Chroma \Ha, [\oiii], RGB \\
2, 3 & Goodhew & Twin APM 6\,in $f/7.9$ refractors & QHYCCD QHY268 Pro M & Astrodon \Ha, [\oiii], LRGB \\
4    & Goodhew & Celestron 14\,in, Starizona Hyperstar $f/1.9$ & QHYCCD QHY268 Pro M & Baader \Ha, [\oiii] \\
5    & Talbot & Stellarvue 6\,in $f/8$ refractor & ZWO ASI6200MM Pro & Astrodon \Ha, Chroma [\oiii] \\
6    & Talbot, Carter & Planewave DeltaRho 350 13.8 in $f/3$ reflector & ZWO ASI461MM Pro & Chroma \Ha, [\oiii], RGB \\
\enddata
\tablecomments{Telescope locations: (1) Sierra Remote Observatories, Auberry, CA, USA; (2, 3, 4)~Fregenal de la Sierra, Spain; (5)~Stark Bayou Observatory, Ocean Springs, MS, USA; (6)~Dark Sky Observatory Collaborative, Fort Davis, TX, USA.  }
\end{deluxetable}

\subsection{LS Pegasi \label{sec:lspeg_imaging_details} }

Images of LS~Peg were obtained by Goodhew between 2020 August~21 and December~5 (Telescopes~2 and~3 in Table~\ref{tab:telescopes}), and by Talbot between 2024 October~25 and December~11 (Telescope~5) and 2024 September~22 and 25 (Telescope~6). Exposure times are summarized in Table~\ref{tab:lspeg_exposures}. The grand-total exposure time was 96.28~hr. Image processing and stacking was done by Goodhew using CCDWare,\footnote{\url{https://ccdware.com/}} CCDStack,\footnote{\url{https://ccdware.com/ccdstack_overview/}} and Adobe Photoshop.\footnote{\url{https://www.adobe.com/products/photoshop.html}} The final image renditions were prepared by Talbot using PixInsight\footnote{\url{https://pixinsight.com}} and Photoshop. 

\begin{deluxetable}{lcccccc}[h]
\tablecaption{Exposure Times [s] on LS Peg \label{tab:lspeg_exposures} }
\tablehead{
\colhead{Telescope}
&\colhead{\Ha}
&\colhead{[\oiii]}
&\colhead{L}
&\colhead{R}
&\colhead{G}
&\colhead{B}
}
\startdata
2 & $71\times1800$ & $1\times1800$ & $15\times300$ & $18\times300$ & $16\times300$ & $15\times300$       \\
3 & $16\times900, 54\times1800$ & $9\times900, 1\times1800 $ & $12\times300$ & $11\times300$ & $9\times300$ & $7\times300$    \\
5 & $29\times1200$ & $8\times1200$ & \dots & $29\times120$ & $30\times120$ & $30\times120$ \\
\noalign{\vskip-0.15in} \\
6 & $20\times480$ & \dots & \dots & \dots & \dots & \dots \\
Total exp.\ [hr] & 78.83 & 5.91 & 2.25 & 3.38 & 3.08 & 2.83 \\
\enddata
\end{deluxetable}

\subsection{ASASSN-V J205457.73+515731.9 \label{sec:asas_imaging_details} }

Images of ASASJ2054 were obtained by Coles from 2024 September~1 to~7, by Goodhew from 2024 August~17 to~28, and by Talbot between 2024 August~18 and September~23. Exposure times are summarized in Table~\ref{tab:asas_exposures}. The grand-total exposure time was 144.33~hr. The rendition shown in our Figures~\ref{fig:asas_image} and \ref{fig:asas_zoom} was created by Goodhew, using CCDWare, CCDStack, PixInsight, and Photoshop. The wide-angle view pressented in Figure~\ref{fig:asas_wideangle} was created by Coles, using PixInsight and Photoshop.

\begin{deluxetable}{lccccc}[h]
\tablecaption{Exposure Times [s] on ASASJ2054 \label{tab:asas_exposures} }
\tablehead{
\colhead{Telescope}
&\colhead{\Ha}
&\colhead{[\oiii]}
&\colhead{R}
&\colhead{G}
&\colhead{B}
}
\startdata
1 & $52\times600 $ & $144\times600$ & $20\times600$ & $12\times600$ & $12\times600$\\
2 & $395\times300$ & \dots & $20\times300$ & $15\times300$& $15\times300$ \\
3 & $234\times300$ & $24\times300 $ & \dots & \dots & \dots \\
4 & $106\times300$ & $131\times300$ & \dots & \dots & \dots \\
5 & $62\times1200 $ & $16\times1200 $ & \dots & \dots & \dots \\
\noalign{\vskip-0.15in} \\
Total exp.\ [hr] & 90.58 & 42.25 & 5.00 & 3.25 & 3.25 \\
\enddata
\end{deluxetable}




\bibliography{PNNisurvey_refs}

\end{document}